\documentclass[aps,prl,twocolumn]{revtex4}
\usepackage{epsfig}
\usepackage{bm}
\begin{document}
\def\B.#1{{\bm #1}}
\def\C.#1{{\cal #1}}
\title{Statistically Preserved Structures and Anomalous Scaling in Turbulent
Active Scalar Advection}
\author{Emily S.C. Ching}
\affiliation{Department of Physics, The Chinese University of Hong Kong,
Sha Tin, Hong Kong}
\author{Yoram Cohen}
\author{Thomas Gilbert}
\author{Itamar Procaccia}
\affiliation{Dept. of Chemical Physics, The Weizmann Institute
of Science, Rehovot 76100, Israel}
\date{\today}
\begin{abstract}
The anomalous scaling of correlation functions in the turbulent statistics
of active scalars (like temperature in turbulent convection)
is understood in terms of an auxiliary passive scalar which is
advected by the same turbulent velocity field. While the odd-order correlation
functions of the active and passive fields differ, we propose that
the even-order
correlation functions are the same to leading order (up to a trivial
multiplicative factor). The leading correlation functions are statistically
preserved structures of the passive scalar decaying problem, and therefore
universality of the scaling exponents of the even-order correlations of
the active scalar is demonstrated.
\end{abstract}
\maketitle
The old riddle of anomalous scaling in {\em generic} turbulent statistics
had been solved recently, albeit in the modest context of 
passive scalar and passive vector advection \cite{01CV,01ABCPV,01CGP}. Motivated
by some recent numerical indications \cite{01CMMV}, we propose
in this Letter that
under generic conditions a similar understanding can be
extended to {\em active} scalar (or vector) advection by
a generic turbulent velocity field. The difference between the
two problems is that
passive fields leave the velocity statistics intact, satisfying
a purely linear problem into which the velocity field is injected
as an advection term, while active fields influence 
the statistics of the velocity field.
This development brings us therefore closer to understanding 
anomalous scaling in the nonlinear velocity problem
itself.

Passive fields are advected in turbulence with typical
dynamics of the form
\begin{equation}
\frac{\partial \phi(\B.r,t)}{\partial t}+\B.u(\B.r,t)\cdot
\B.\nabla \phi(\B.r,t)=\kappa \nabla^2 \phi(\B.r,t)+f(\B.r,t) \ . 
\label{passive}
\end{equation}
Here the field $\phi(\B.r,t)$ can be a passive scalar $c(\B.r,t)$
or passive vector $\B.v(\B.r,t)$. $\kappa$ is the diffusivity
and $f(\B.r,t)$ is a white random force of zero mean
with compact support in $\B.k$-space, acting on the largest
scales of the order of the outer scale $L$ only. The advecting
velocity field is a generic turbulent field at high Reynolds
number with an extended scaling range, typically exhibiting 
anomalous scaling. We use the word ``generic'' here to distinguish
from model velocity fields with $\delta$-function temporal 
correlations \cite{01FGV}. ``Anomalous scaling'' means that multi-point 
correlation functions are homogeneous functions of their
arguments, with exponents that cannot be guessed from dimensional
analysis. Thus for example the field $\phi(\B.r,t)$ has 
simultaneous multi-point correlation functions
\begin{equation}
F^{(n)}(\B.r_1,\B.r_2,\!\cdots\!, \B.r_n)\equiv
\langle \phi(\B.r_1,t)\phi(\B.r_2,t)\!\cdots\!
\phi(\B.r_n,t)\rangle_f \ ,
\end{equation}
where pointed brackets with subscript $f$ refer to 
averaging over the statistics
of the advecting velocity field {\em and} of the forcing.
Anomalous scaling means that 
\begin{equation}
F^{(n)}(\lambda \B.r_1, \cdots, \lambda\B.r_n)
=\lambda^{\zeta_n} F^{(n)}(\B.r_1,\cdots, \B.r_n) \ ,
\end{equation}
with $\zeta_n$ having a non-trivial dependence on $n$.

In recent work \cite{01ABCPV,01CGP} it was clarified why and how passive fields
exhibit anomalous scaling. The key is to consider a problem 
associated with Eq. (\ref{passive}) which is the {\em decaying
problem} in which the forcing $f(\B.r,t)$ is put to zero. 
The problem becomes then a linear initial value problem,
$\frac {\partial \phi}{\partial t} =\C.L \phi$, 
with a formal solution
\begin{equation}
\phi(\B.r,t) = \int d\B.r' \B.R(\B.r,\B.r',t) \phi(\B.r',0) \ , \label{oper}
\end{equation} 
with the operator $\B.R\equiv T^+ \exp\left({\int_0^t ds \C.L(s)}\right)$,
and $T^+$ being the time ordering operator.
Define next the now {\em time dependent} correlation
functions of the decaying problem:
\begin{equation}
C^{(n)}(\B.r_1,\cdots ,\B.r_n,t)\equiv
\langle \phi(\B.r_1,t)\cdots
\phi(\B.r_n,t)\rangle \ .
\end{equation}
Pointed brackets without subscript $f$ refer to the decaying
object in which averaging is with respect to realizations of
the velocity field only. As a result of Eq. (\ref{oper}) the decaying
correlation functions are developed by a propagator 
$\C.P^{(n)}_{\underline{\B.r}|\underline{\B.\rho}}$, (with
$\underline {\B.r}\equiv \B.r_1,\B.r_2,\!\cdots\!,\B.r_n$):
\begin{equation}
C^{(n)}(\B.r_1,\!\cdots\!, \B.r_n,t)=\int\!\! d\underline{\B.\rho}
 \C.P^{(n)}_{\underline{\B.r}|\underline{\B.\rho}}(t)~
C^{(n)}(\B.\rho_1,\!\cdots\!, \B.\rho_n,0) \ .
\end{equation}

In writing this equation we made explicit use of the fact that
the {\em initial} distribution of the passive field $\phi(\B.r,0)$
is statistically independent of the advecting velocity field. Thus
the operator $\C.P^{(n)}_{\underline{\B.r}|\underline{\B.\rho}}$ 
can be written explicitly
\begin{equation}
\C.P^{(n)}_{\underline{\B.r}|\underline{\B.\rho}}(t)\equiv
\langle  \B.R(\B.r_1,\B.\rho_1,t) \B.R(\B.r_2,\B.\rho_2,t)\cdots 
\B.R(\B.r_n,\B.\rho_n,t)\rangle \ .
\end{equation}

The key finding \cite{01ABCPV,01CGP} is that the operator 
$\C.P^{(n)}_{\underline{\B.r}|
\underline{\B.\rho}}$ possesses {\em left} eigenfunctions of eigenvalue
1, i.e. there exist functions $Z^{(n)}(\B.r_1,\B.r_2\cdots \B.r_n)$
satisfying
\begin{equation}
Z^{(n)}(\B.r_1,\cdots, \B.r_n)=\int d\underline {\B.\rho}
\C.P^{(n)}_{\underline{\B.r}|\underline{\B.\rho}}(t)
Z^{(n)}(\B.\rho_1,\cdots, \B.\rho_n) \ .
\end{equation}
The functions $Z^{(n)}$ are referred to as ``statistically preserved
structures'', being invariant to the dynamics, even though
{\em the operator is strongly time dependent and decaying}. How to form from these
functions infinitely many conserved variables in the decaying
problem was shown in \cite{01ABCPV}. The functions
$Z^{(n)}(\underline{\B.r})$ are homogeneous functions of
their arguments, with anomalous scaling exponents $\zeta_n$.
More importantly,
it was shown that the correlation functions of the forced case,
$F^{(n)}(\underline{\B.r})$, have exactly the same scaling exponents
as $Z^{(n)}(\underline{\B.r})$ \cite{01CGP}. In the scaling sense
\begin{equation}
F^{(n)}(\underline{\B.r})\sim Z^{(n)}(\underline{\B.r}) \ . \label{scalesame}
\end{equation}
This is how anomalous scaling in passive fields is understood.

Consider next the dynamics of an active field. For concreteness
we will think about temperature in turbulent Boussinesq convection, but the
ideas are immediately generalized to any other type of active
field. The equation of motion is formally identical to Eq.~(\ref{passive}):
\begin{equation}
\frac{\partial T(\B.r,t)}{\partial t}\!+\!\B.u(\B.r,t)\cdot
\B.\nabla T(\B.r,t)\!=\!\kappa \nabla^2 T(\B.r,t)\!+\!f(\B.r,t)  , 
\label{active}
\end{equation}
but now the velocity field is affected by the temperature. For
an incompressible fluid of unit density \cite{71MY},
\begin{equation}
\frac{\partial \B.u}{\partial t}+\B.u\cdot
\B.\nabla \B.u=-\B.\nabla p+\nu \nabla^2 \B.u
+\alpha g T\hat z \ . 
\label{Boussinesq}
\end{equation}
Here $p$, $\nu$, $\alpha$, $g$ and $\hat z$ are the pressure, 
kinematic viscosity, thermal
expansion coefficient, acceleration due to gravity and a unit
vector in the upward direction respectively. The 
appearance of $T$ in the equation for $\B.u$ is crucial,
and changes the scaling exponents of $\B.u$ from Kolmogorov 
(approximately) to Bolgiano (approximately) \cite{71MY}. 
It makes no sense now to consider the decaying problem
for $T$; as this field decays, the statistics of the velocity field
changes, and there is very little to say. On the other 
hand, we can learn a great deal from considering the forced
solutions. We note that both the passive and the active forced
equations can be solved in the same way: 
\begin{eqnarray}
T(\B.r,t) = &&\int d\B.r' \B.R(\B.r,\B.r',t)~T(\B.r',0)\nonumber\\+
 &&\int d\B.r'\int_0^t d\tau \B.R(\B.r,\B.r',t-\tau)f(\B.r',\tau)\ .
\end{eqnarray}
with a similar expression for the passive scalar. For 
$t\to \infty$, the first term decays to zero, but the second
term exists. Next  we encounter the difference between the 
active and passive case. Computing the average of $T$, or any
other odd moment, we get a finite result:
\begin{equation}
\langle T(\B.r,t)\rangle  = \int d\B.r'\int_0^t d\tau 
\langle \B.R(\B.r,\B.r',t-\tau)f(\B.r',\tau)\rangle \ . \label{mean}
\end{equation}
We cannot decorrelate the random forcing $f$ from the 
operator $\B.R$ which involves the velocity field. The forcing $f$
is correlated with $T$, which is itself correlated, via Eq.(\ref{Boussinesq}),
with $\B.u$,
and therefore the field $T$ has a first and higher odd moments: the
probability density function (pdf) of $T$ will be asymmetric. 
For the passive case, the equivalent Eq. (\ref{mean})
can be simplified by decorrelating $f$ from $\B.R$. As the odd moments
of $f$ vanish, the passive scalar has zero odd moments, and
its pdf is symmetric. This difference is fundamental.

We now argue that all the {\em even} moments, and
all even correlation functions of the active and passive fields
may be identical to leading scaling order. Consider for example
the second order correlation function $F^{(2)}(\B.r_1,\B.r_2)$
of the active scalar:
\begin{eqnarray}
&&F^{(2)}(\B.r_1,\B.r_2)=\int d\B.r' d\B.r'' \int_0^t d\tau' d\tau''\nonumber\\
&&\langle \B.R(\B.r,\B.r',t-\tau')f(\B.r',\tau')\B.R(\B.r,\B.r'',t-\tau'')
f(\B.r'',\tau'')\rangle \ . \label{second}
\end{eqnarray}
As before, in the case of the active scalar the forces cannot be
decorrelated from the operators. For the passive scalar this can
be done with impunity, leading to the correct form of the correlation
function, i.e.
\begin{eqnarray}
\langle \phi(\B.r_1)\phi(\B.r_2)\rangle_f& =& \int d\B.\rho_1d\B.\rho_2 
\int_0^t d\tau \C.P^{(2)}_{\B.r_1,\B.r_2|\B.\rho_1\B.\rho_2}(t-\tau)\nonumber\\ 
&\times& \langle f(\B.\rho_1) f(\B.\rho_2)\rangle \ . \label{spassive}
\end{eqnarray}
It was shown in \cite{01CGP} how this expression indeed yields
the forced correlation function in agreement with Eq.(\ref{scalesame}).
We observe however that any correlation function $\langle A B\rangle$
can be written in terms of its Gaussian decomposition and the
``connected part'':
$\langle AB\rangle = \langle A\rangle \langle B\rangle +
\langle \tilde A \tilde  B\rangle $,
where $\tilde A\equiv A -\langle A\rangle$. Identify now $A$ with
the product of operators $\B.R$ in Eq. (\ref{second}), and
$B$ with the product of forcing, and realize that Eq.(\ref{spassive})
is the Gaussian decomposition contribution to Eq. (\ref{second}).
So the solution (\ref{spassive}) exists as one contribution to (\ref{second}),
the other being what remains of the correlation, i.e. the part in which
the forcing cannot be decorrelated from the operator. A similar
phenomenon occurs for {\em all} the even order correlation
functions. But we know
from the theory of the passive scalar that the solution (\ref{spassive})
and the equivalent ones for all the even-order  correlation functions
are identified with statistically preserved structures whose scaling
is anomalous \cite{01CGP}. We propose that the contributions
to the even-order correlation of the active scalar which are identical
to the even-order  correlations of the passive scalar are leading, and
are therefore expected to be realized in experiments and simulations.
The expectation is that the connected contributions, that are 
purely ``forced''
solutions, are either subleading or have the same exponents. This
is a crucial conjecture that needs to be tested against examples. In 
\cite{01CMMV} it was shown that this is correct for a particular
4th order correlation function of the active and passive fields.

To exemplify the generality and plausibility of these ideas
we examine in detail a model of active and passive scalar
for which the statistical object can be computed to high
accuracy. We consider a variant of the shell model studied
in ref. \cite{Brand}:
\begin{widetext}
\begin{eqnarray}
\frac{\partial T_n}{\partial t}&=&\tilde ak_n(u_{n-1}T_{n-1}-hu_nT_{n+1})
+\tilde b k_n(u_nT_{n-1}-hu_{n+1}T_{n+1})-\kappa k_n^2 T_n+f\delta_{n,0}\\
\frac{\partial u_n}{\partial t}&=&ak_n(u_{n-1}^2-hu_nu_{n+1})
+bk_n(u_nu_{n-1}-hu^2_{n+1})-\nu k_n^2 u_n+T_n\\
\frac{\partial C_n}{\partial t}&=&\tilde ak_n(u_{n-1}C_{n-1}-hu_nC_{n+1})
+\tilde bk_n(u_n C_{n-1}-hu_{n+1}C_{n+1})-\kappa k_n^2 C_n+f\delta_{n,0} \ .
\end{eqnarray}
\end{widetext}
In this model $n$ stands for index of a shell of k-vector $k_n=k_0h^n$,
with $n=0,1,\cdots, N-1$. We take $h=2$, and the parameters used
in the simulation are $a=0.01$, $\tilde a=\tilde b=b=1$,
$k_0=1$, $\kappa=\nu=5\times 10^{-4}$. The number of shells
is $N=30$. 

\begin{figure}
\centering
\includegraphics[width=.33\textwidth]{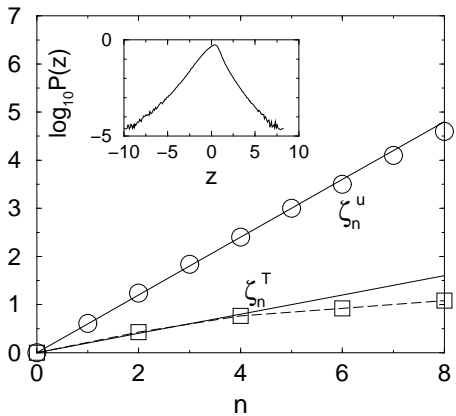}
\caption{The scaling exponents for the velocity field (circles)
and the active scalar field (squares).
The solid lines are respectively $3n/5$ and $n/5$ for the velocity
and the active scalar fields. Shown in the inset is 
the pdf of $z=\tilde u_n/\langle \tilde u_n^2 \rangle^{1/2}$ 
at shell $n=14$.}
\label{scexp}
\end{figure}

Without the coupling to $T_n$, the velocity equation has
an inviscid unstable Kolmogorov fixed point, $u_n\sim k_n^{-1/3}$.
This is changed by the coupling \cite{Brand}, and the system of equations
for $T_n$ and $u_n$ exhibits an inviscid unstable Bolgiano
fixed point, $u_n\sim k_n^{-3/5}$, $T_n\sim k_n^{-1/5}$.
The chaotic dynamics renders the statistics of the velocity strongly
non-Gaussian, cf. inset in Fig \ref{scexp}. The exponents for the
active scalar are markedly anomalous, whereas for the velocity 
appear closer to normal, see Fig. \ref{scexp}. 

\begin{figure}
\centering
\includegraphics[width=.33\textwidth]{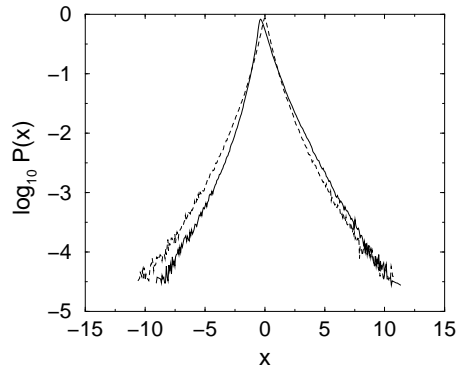}
\caption{The pdf's of the active (solid) and passive (dashed) 
scalars at shell $n=14$. Note that the pdf of 
the active scalar is asymmetric.}
\label{pdf}
\end{figure}

\begin{figure}
\centering
\includegraphics[width=.33\textwidth]{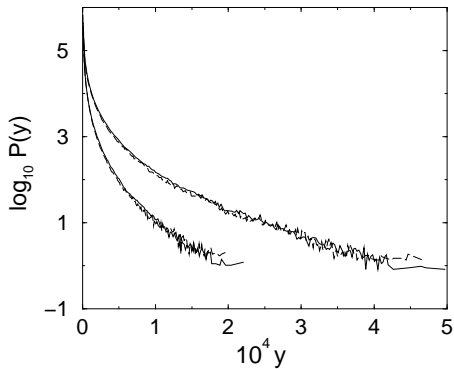}
\caption{
The pdf's of $y$ where $y=\tilde T_n^2$ (solid) 
or $y=\beta C_n^2$ (dashed) at shells $n=9$ and 14.}
\label{pdfsq}
\end{figure}

\begin{figure}
\centering
\includegraphics[width=.33\textwidth]{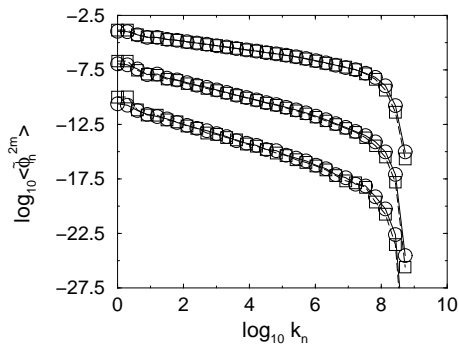}
\caption{
The even-order structure functions $\langle \tilde T_n^{2m} \rangle$ (circles)
and $\langle \beta^m C_n^{2m} \rangle$ (squares), with 
$m=1$, 2 and 3, from top to bottom.}
\label{structure}
\end{figure}

\begin{figure}
\centering
\includegraphics[width=.33\textwidth]{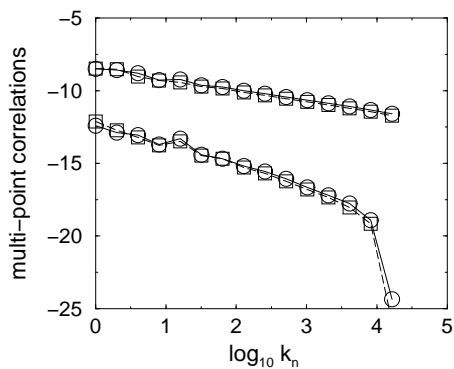}
\caption{Upper:
$\log_{10}\langle \tilde T_n^2\tilde T_{n+5}^2 \rangle$ (circles) and  
$\log_{10}\langle \beta^2 C_n^2C_{n+5}^2 \rangle$ (squares). Lower:
$\log_{10} \langle \tilde T_n^2\tilde T_{n+5}^2\tilde T_{2n}^2 \rangle$ (circles) 
and $\log_{10} \langle \beta^3C_n^2C_{n+5}^2C_{2n}^2 \rangle$ (squares).}
\label{multi}
\end{figure}

To demonstrate the first point of the fundamental 
difference between the active and passive fields, we show in Fig. \ref{pdf}
the pdf's of $x=\tilde \phi_n/\langle \tilde \phi_n^2\rangle^{1/2}$ 
where $\tilde \phi$ is $\tilde T_n$ or $C_n$,
for $n=14$. One clearly sees the symmetry of the pdf
of the passive scalar, in contradistinction to the asymmetry
of the pdf of the active scalar. This is typical to all $n$ 
in the inertial range. This is a demonstration of the
discussion after Eq.(\ref{mean}). For the passive scalar the
odd moments vanish, whereas for the active scalar they all exist.
The situation is altogether different for the statistics of
even moments. To demonstrate the difference we plot in Fig. \ref{pdfsq}
the (typical) pdf of $\tilde T_n^2$ and $C_n^2$ for $n=9$ and 14. In plotting
we realize that the passive scalar is defined up to a constant,
so for the passive scalar the pdf is plotted for the rescaled
variable $\beta C^2_n$, where $\beta=\langle \tilde T _n^2\rangle /
\langle C_n^2\rangle \approx 0.6327$. Note that there is only one
numerical freedom $\beta$, constant for all $n$ in the inertial range.
We find very close agreement of all the pdf's in the
inertial range.
The identity of the pdf's of $\tilde T_n^2$ and $C_n^2$ translates
automatically to the identity (up to a constant $\beta^m$) of the even-order 
structure functions $S^{(2m)}(k_n)\equiv \langle \tilde \phi_n^{2m}\rangle$,
where $\tilde \phi_n = \tilde T_n$ or $C_n$. This is demonstrated 
in Fig. \ref{structure}.
We see that the 2nd, 4th and 6th-order
structure functions are barely distinguishable, with the
same scaling exponents in the inertial range.
Finally, the identity of the statistics of the squares of the passive and
active scalars transcends structure functions. We consider
next multi-point correlation functions, and in Fig. \ref{multi}
compare the correlation functions $\langle \tilde T_n^2
\tilde T^2_{n+5}\rangle$
and $\langle \tilde T^2_n \tilde T^2_{n+5}\tilde T^2_{2n}
\rangle$ to their passive counterparts. 
The conclusion is that again the multi-point correlation
functions are indistinguishable once the passive
ones are rescaled by $\beta^q$ where $q$ is the over-all
order of the correlation function.
In summary, we offered a discussion and detailed numerical evidence to
support the conjecture that under generic conditions the even-order
correlation functions of active scalars can be understood via the
emerging theory of Statistically Preserved Structures of the passive
scalar counterpart. We have by now two examples, together with
\cite{01CMMV}, which share however a similar coupling of the
active scalar to the velocity field. The importance of this conjecture
warrants considerable further work to uncover its provisos and delineate
its generality.
\begin{acknowledgments}
This work had been supported in part by the
Research Grants Council of Hong Kong SAR, China
(CUHK 4119/98P and CUHK 4286/00P), the European Commission
under a TMR grant, the German Israeli Foundation, and the
Naftali and Anna Backenroth-Bronicki Fund for Research in
Chaos and Complexity. TG thanks the Israeli Council for Higher Education and the Feinberg 
postdoctoral Fellowships program at the WIS for financial support.
\end{acknowledgments}

\end{document}